\documentclass[floats,twocolumn,prl,aps]{revtex4}

\usepackage{graphicx}

\newcommand\beq{\begin{equation}}
\newcommand\eeq{\end{equation}}
\newcommand\bea{\begin{eqnarray}}
\newcommand\eea{\end{eqnarray}}
\newcommand{\nonum}{\nonumber}

\begin{document}

\title{\bf Excitations of a low-dimensional dimerized spin ladder 
under a magnetic field.}

\author{\bf Sujit Sarkar}
\address{\it PornaPrajna Institute of Scientific Research,
4 Sadashivanagar, Bangalore 5600 80 and   
 Department of Condensed Matter Physics,
Weizmann Institute of Science, Rehovot-76100, Israel}

\date{\today}

\begin{abstract}
Here we study the dimerized spin ladder with nearest-neighbor ($J_1$) and next-nearest-neighbor 
($J_2$) anti-ferromagnetic interaction under a magnetic field. We predict the existence
of different magnetization plateaus for the presence of spin-Peierls interaction on both $J_1$ and $J_2$. Magnetization plateau
at $m=0$ for $J_1$ dimerization is spontaneous due to XY interaction, but it is absent for
$J_2$ dimerization, only intrinsic umklapp term leads to plateau (spin gap) state for some specific
values of XXZ anisotropy ($\Delta$) and $J_2$.
Here we predict a saturation plateau which
is the classical phase of the system. There are some numerical
support of our analytical approach already existing
in the literature. The transition from commensurate gapped phase to
incommensurate Luttinger liquid phase is the Mott-$\delta$ type of transition.
\vskip .4 true cm
\noindent PACS numbers: ~75.10.Jm-Quantized Spin Models 
, 75.40.Cx-Static Properties (order parameters, static susceptibilities,
heat capacities, critical exponents etc)
, 75.45.+j-Macroscopic Quantum Phenomena in Magnetic Systems.

\end{abstract}

\maketitle


\section{ 1. Introduction}

One-dimensional and quasi-one-dimensional 
quantum spin systems have been studied extensively for last few decades,
due to many unusual and interesting 
findings from both experimental and theoretical sides.
The four observations which make the low-dimensional spin systems 
particularly interesting are, (i) Haldane's conjecture for one-dimensional 
anti-ferromagnetic spin systems \cite{hal}, (ii) the discovery of 
high-temperature superconductivity and its magnetic properties at low doping 
\cite{bed}, (iii) the discovery of ladder materials \cite{joh}, and (iv) response of the
spin-ladder/chain systems under a magnetic field 
(gapped excitations, magnetization plateau, magnetization cusp,
and first order phase transition) \cite{chit,oshi,tot,shi,tone,kole,cab,mutt,oku,sato}.
It has been observed
that spin-1/2 ladder systems with the railroad geometry with an even number of legs are gapped,
while systems with an odd number of chains have gapless excitations \cite{dag,cha}.
However, the frustrated zigzag ladder
shows gapless spin liquid state or the gapped dimer state, depending
on the ratio of the exchanges of the rungs to the chains \cite{egg}.

However, all these above mentioned studies are in the absence of magnetic field. The situation
become very interesting in presence of magnetic field. Then it is possible 
for an integer spin chain to be gapless with partial magnetization and a half-odd-integer
spin chain to show a gap above the ground state for appropriate values of
magnetic field. It has been shown by different groups \cite{oshi,tot,shi,tone,kole}
that the magnetization of some systems
can exhibit plateaus at certain nonzero values for some finite ranges of the magnetic field.
The basic criteria for the appearance of magnetization plateau can be understood from the 
extension of Lieb-Schultz-Mattis theorem under a magnetic field. This implies that translationally
invariant spin chains in an applied field can be gapped without breaking translation symmetry
under the condition $S-m=$ integer, where $S$ is the spin
and $m$ is the magnetization state of the chain. In this gapped phase, magnetization plateau occurs for quantized
values of m. Fractional quantization can also occur, if accompanied by spontaneous breaking of
translational symmetry. Fractional quantization can be understood from the $S-m=$ non integer
condition. In this situation system is either in the gapless low lying states or the degenerate 
ground state with spontaneous translational symmetry breaking in the thermodynamic limit.
Suppose a system is in the spontaneously translational symmetry breaking state, the ground state
should be degenerate (say q fold). In this situation ground state have q different eigenvalues of
translational operator , the degenerate states can be related to the spontaneous breaking of
translational symmetry to period of q sites in the thermodynamic limit. These conditions for the
appearance of plateau are the necessary but not the sufficient conditions. The nature and the
occurrence of the plateau also depend on the nature of the interaction present
in the system \cite{cab,mutt}. 
Apart from this interesting plateau formation,the presence of magnetic field has also other
interesting effects on spin systems \cite{jap,pok,park,oku}. 

In our present study, we will study the effect of magnetic field on the
dimerized (both rung and leg), spin ladder. Here we predict the presence of
three fractionally quantized magnetization plateaus ($m=0,~1/4,~1/2$)
whereas the other fractionally quantized plateaus are absent. We mainly
stress on the study of leg-dimerization. 
\cite{tot}.
Fig. 1 shows the zig-zag ladder with dimerized legs
(different chain exchange). Rung dimerization is the dimerization in $J_1$, which has not been shown
in the figure.
\begin{figure} 
\includegraphics*[scale=.5,angle=-90]{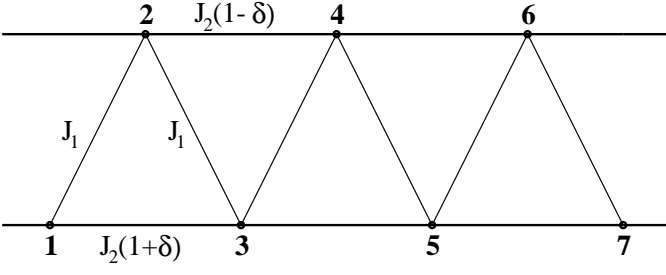}
\caption{Schematic diagram of a spin ladder with different chain exchanges.} 
\label{fig1}
\end{figure}
We compare our calculation and results at
each and every steps with rung dimerization with a marginally  
new result in rung dimerization regarding the
XXZ anisotropy of NN interaction. In this work our approach is completely analytic,
we use abelian
bosonization and renormalization group method to study these model Hamiltonians. We compare
our results with the available numerical results. Sections of this article, are the following. 
Section (2). 
Hamiltonians and the general formulation of analytical calculations, with the analysis of
different magnetization plateau. Section (3) is devoted for discussions and conclusions.  
 
\section{ 2. Model Hamiltonians and Continuum Field Theoretical Study:}

In this section we will present all general derivations of this paper, different
subsections for different magnetization plateaus, are the special limit of these general
derivations.  
The model Hamiltonian for $J_2$ and $J_1$ dimerization are respectively,
\bea
H_A &=& J_1 \sum_{n}~( {{S}_n}(x)  {{S}_{n+1}(x)}~+~
{{S}_n}(y)  {{S}_{n+1}(y)} ~, \nonum \\
&&+ \Delta {{S}_n}(z)  {{S}_{n+1}(z)} )+ 
 J_2 \sum_{n} (1 - \delta (-1)^n) \vec{S}_n \cdot \vec{S}_{n+2} \nonum \\
&& -{g {\mu}_B H} \sum_{n} {S_n} (z)
\label{hamB}
\eea 
\bea
H_B &=& J_1 \sum_{n} (1 - \delta (-1)^n)~( {{S}_n}(x) {{S}_{n+1}(x)}~+~
{{S}_n}(y) {{S}_{n+1}(y)} , \nonum \\
& & +\Delta~~ {{S}_n}(z) {{S}_{n+1}(z)}~~) + 
J_2 \sum_{n} \vec{S}_n \cdot \vec{S}_{n+2} \nonum \\
& & -{g {\mu}_B H} \sum_{n} {S_n} (z)
\label{hamA}
\eea

Where n is the site index, x, y, and z are components of spin. 
One can also recast the zigzag ladder as a linear chain with
NN and NNN exchange interactions \cite{sar1,gia2}.
$J_1$ and $J_2$ are the nearest-neighbor and next-nearest-neighbor
exchange coupling between spins, ${J_1}, {J_2}~\geq 0$, 
$\Delta$ is z component anisotropy of NN exchange interaction. $\delta$ is dimerization
strength, it appear as a parameter in our Hamiltonian, the basic origin of this 
parameter is the spin-phonon interaction.  
$H$ is the externally applied magnetic field in the z direction. One can express
spin chain systems to a spinless fermions systems through 
the application of Jordan-Wigner transformation. In Jordan-Wigner transformation
the relation between the spin and the electron creation and annihilation operators are
\bea
S_n^z & = & \psi_n^{\dagger} \psi_n - 1/2 ~, \nonum \\
S_n^- & = &  \psi_n ~\exp [i \pi \sum_{j=-\infty}^{n-1} n_j]~, \nonum \\
S_n^+ & = & \psi_n^{\dagger} ~\exp [-i \pi \sum_{j=-\infty}^{n-1} n_j]~,
\label{jor}
\eea
where $n_j = \psi_j^{\dagger} \psi_j$ is the fermion number at site $j$. 
\bea
{H}_{A1}~&=& - \frac{J_1}{2} ~\sum_n ~(\psi_{n+1}^{\dagger} \psi_n + 
\psi_n^{\dagger} \psi_{n+1}) \nonum \\
&& + J_1 \Delta \sum_n (\psi_n^{\dagger} \psi_n - 1/2) (\psi_{n+1}^{\dagger} 
\psi_{n+1} - 1/2) ~,\nonum \\
&&  -g {{\mu}_B} H \sum_{n} (\psi_n^{\dagger} \psi_n - 1/2).
\label{ha1}
\eea
\bea
H_{A2}~&=& J_2 ~\sum_n ~( \psi_{n+2}^{\dagger} \psi_n + {\rm h.c.})
(\psi_{n+1}^{\dagger} \psi_{n+1} - 1/2) \nonum \\
&& +~ J_2 ~\sum_n ~(\psi_n^{\dagger} \psi_n - 1/2)
(\psi_{n+2}^{\dagger} \psi_{n+2} - 1/2) .
\label{hA3}
\eea
\bea
H_{A3}~&=& -~J_2 \delta \sum_n (-1)^n ( \psi_{n+2}^{\dagger} \psi_n + {\rm h.c.})
(\psi_{n+1}^{\dagger} \psi_{n+1} - 1/2) \nonum \\
&& - ~ J_2 \delta  \sum_n (-1)^n ~(\psi_n^{\dagger} \psi_n - 1/2)
(\psi_{n+2}^{\dagger} \psi_{n+2} - 1/2).
\label{hb3}
\eea
There is a difference between the first term of Eq.4 with the first
term of Eq.5 and Eq.6, in Eq.4 first term present the hopping 
whereas the first term in Eq.5 and Eq.6 are presenting the four 
fermionic interaction. This difference arises due to the presence
of an extra factor $e^{- i \pi {n}_{j+1}}$ in the string of
Jordan-Wigner transformation for NNN exchange interactions.

Similarly one can also recast the spin-chain systems with $J_1$ 
dimerization into the spinless fermions. The Hamiltonians are converted as 
follows:
$H_{B1}~=~H_{A1}$, $H_{B2}~=~H_{A2}$ and 
\bea
{H}_{B3}~&=& \frac{J_1}{2} ~\delta \sum_n (-1)^n ~(\psi_{n+1}^{\dagger} \psi_n + 
\psi_n^{\dagger} \psi_{n+1}) \nonum \\
&& -~J_1 \delta \Delta \sum_n ~(-1)^n (\psi_n^{\dagger} \psi_n - 1/2) (\psi_{n+1}^{\dagger} 
\psi_{n+1} - 1/2),
\label{ha2}
\eea
In order to study the continuum field theory of these Hamiltonians, we recast the spinless
fermions operators in terms of field operators by this relation. 
\beq
{\psi}(x)~=~~[e^{i k_F x} ~ {\psi}_{R}(x)~+~e^{-i k_F x} ~ {\psi}_{L}(x)]
\eeq
where ${\psi}_{R} (x)$ and ${\psi}_{L}(x) $ are describe the second-quantized fields of right- and 
left-moving fermions respectively. The spin-Peierls compound, in the 
absence of magnetic field, is Generally at half-filling. 
In the absence of magnetic field ($H=0$), $k_F~=~\pm {\pi}/2$,
but we are interested to study the systems in presence of magnetic field, so we keep Fermi 
momentum as arbitrary $k_F$.
One can simply absorb the finite magnetization in a shift of field $\phi$ by
$\phi~=~{\tilde {\phi} - \pi m x}$, where $m~=<S_z>$ .
In presence of
magnetic field Fermi momentum and magnetization ($m$) are related by this relation,
$k_F~=~~\frac{\pi}{2} ( 1~-~2 m)$ \cite{tot,gia2}.
We want to express the fermionic fields in terms of bosonic field by this relation 
\beq
{{\psi}_{r}} (x)~=~~\frac{U_r}{\sqrt{2 \pi \alpha}}~~e^{-i ~(r \phi (x)~-~ \theta (x))} 
\eeq
$r$ is denoting the chirality of the fermionic fields,
 right (1) or left movers (-1).
The operators $U_r$ are operators that commute with the bosonic field. $U_r$ of different species
commute and $U_r$ of the same species anticommute. $\phi$ field corresponds to the 
quantum fluctuations (bosonic) of spin and $\theta$ is the dual field of $\phi$. They are
related by this relation $ {\phi}_{R}~=~~ \theta ~-~ \phi$ and  $ {\phi}_{L}~=~~ \theta ~+~ \phi$.

Using the standard machinery of continuum field theory \cite{gog,gia2,sumo,tot}, 
we finally get the bosonized Hamiltonians
as 
\bea
H_{0}~&=&~v_0 \int_{o}^{L} \frac{dx}{2 \pi} \{ {\pi}^2 : {\Pi}^2 : ~+~ :[ {\partial}_{x} \phi (x) ]^2 : \nonum\\
&&~+~ \frac{g_1}{{\pi}^2 }\int ~ dx~:[ {\partial}_{x} {{\phi}_L} (x) ]^2  :
+ :[ {\partial}_{x} {{\phi}_R} (x) ]^2 : \nonum\\
&&~+~ \frac{g_2}{{\pi}^2 }\int ~ dx~ 
 ({\partial}_{x} {{\phi}_L} (x) ) ({\partial}_{x} {{\phi}_R} (x) ) 
\label{bos1}
\eea
$H_{0}$ is the gapless Tomonoga-Luttinger liquid part of the Hamiltonian
with $v_0~=sink_F$. The analytical expressions for $g_1$ and $g_2$
(related with the forward scattering of fermionic field) are the following. 
\beq
g_1~=~2 (\Delta ~-~ 2 J_2) ~{sin}^2 k_F ~+~ 2~J_2 ~ sin 2 k_F (\pi~+~sin 2 k_F) \nonum\\ 
\eeq
\beq
g_2~=~4 (\Delta ~-~ 2 J_2) ~{sin}^2 k_F ~+~ 4~J_2 ~ {sin}^{2} {2 k_F } \nonum\\ 
\eeq
The expression for $g_1$ and $g_2$ are the same of Ref. \cite{hal2}
and Ref. \cite{tot}
at $m=0$ ($k_F=\pi/2$). 

Analytical expressions for 
different exchange interactions of Hamiltonian, $H_A$, are the following.

\beq
H_{J2C1}~=~ \frac{J_2}{2 {\pi}^2 {\alpha}^2}
\int dx :cos[4 \sqrt{K}\phi (x) ~-~ 4 k_F x~-~ 4 k_F a]: .
\eeq
\beq
H_{J2C2}~=~\frac{J_1 \Delta}{2 {\pi}^2 {\alpha}^2}
\int dx :cos[4 \sqrt{K}\phi (x) ~+~ 4 k_F x~-~2 k_F a]: .
\eeq
\beq
H_{J2C3}=  \frac{J_2 \delta}{2 {\pi}^2 {\alpha}^2}
 \int dx :cos[(\pi- 4 k_F)x~+~4 \sqrt{K}~\phi (x) ~-~4 k_F a]:.
\eeq
Eq.13 and Eq.14 are presenting the umklapp scattering term
from the NN and NNN anti-ferromagnetic exchange interaction, Eq.15
is appearing due to the presence of dimerized interaction.
Analytical expressions for $K$ is the following.
\beq
K = {\Large[}\frac{1~-~(8/\pi)~J_2 ~{sin}^2 k_F ~+~ 4 J_2~cos k_F}
      {1~+~(4/\pi) \Delta sin k_F~+~ 4 J_2 ~ cos k_F ( 1~+~2/\pi  sin {2 k_F }) }{\Large]}^{1/2}.
\eeq
$v_0$ and $K$ are the two Luttinger liquid parameters.
The expression of K for $J_2 ~=0$ and $m=0$ is the same
as in Ref. \cite{gia2}.
During this derivation we have used the following relations: 
${\rho}_{R/L} ~=~\frac{-1}{\pi}{\partial}_{x} {\phi}_{R/L} (x)$ and
$[{\phi} (x) ~,~\Pi (x) ]~= i \delta (x-x')$, where
$\Pi (x)~=\frac{1}{\pi} \nabla \theta (x)$,
is the canonically conjugate momentum.
We have also used the following equations,
\beq
S^{z} (x)~= ~a ~[ ~\rho (x)~+~ (-1)^j ~ M(x) ~] ~.
\label{basic1}
\eeq
The bosonized expressions for $\rho$ and $M$ are given by
\bea
\rho (x) &=& ~-~ \frac{1}{\sqrt \pi} ~\partial_x \phi (x) ~, \nonum \\
M(x) &=& ~\frac{1}{\pi a} ~\cos (2  \phi (x)) ~.
\label{basic2}
\eea
Similarly one can calculate the analytical expressions for $J_1$ dimerization.
Totsuka \cite{tot} has already calculated the bosonized expressions for $J_1$ dimerization only
by using the symmetric convention of chiral fermions 
($,{\psi}_L ~\sim e^{-2 i {\phi}_L}$,
${\psi}_R ~\sim e^{-2 i {\phi}_R}$),
and has expressed the sine-Gordon coupling, in terms of
dual field ($\theta$). Here we have expressed our all expressions in terms of bare phase field
($\phi$), by
using the conventional practice of continuum field theory \cite{gog,gia2,sumo}. 
During these derivations we assume
that $J_1 \gg J_2 ,~ \delta$. $J_2$ is in the unit of $J_1$. Here we neglect the
higher order
of $a$ than $a^2$.

\subsection{ 2.1 Calculations and Results for $m=0$ Magnetization Plateau:}

At first we discuss $m=0$ magnetization plateau, it corresponds $k_F ~=~\pm~ \pi/2$.
Here we study both the effect
of XXZ anisotropy ($\Delta$) and the spin-Peierls dimerization ($\delta$).
The effective Hamiltonian for $J_2$ dimerization become, 
\beq
H_A = H_0 ~+~( \frac{J_2 ~-~\Delta}{2 {\pi}^2 {\alpha}^2} )~\int dx :cos[4 \sqrt{K}~\phi (x)]: .
\label{ha0}
\eeq
In this effective Hamiltonian (Eq.19), there is no contribution
from dimerized
interaction
due to the oscillatory nature of the integrand and it leads to a vanishing
contribution but the contribution of dimerized potential is present
in the NN exchange interaction. 
Similarly the effective Hamiltonian for $J_1$ dimerization become, 
\bea
H_B & = & H_0 ~+~ ( \frac{J_2 ~-~\Delta}{2 {\pi}^2 {\alpha}^2} )~\int dx :cos[4 \sqrt{K}~\phi (x)]: \nonum\\
&& ~+~\frac{\delta}{2 {\pi}^2 {\alpha}^2}\int dx :cos[(2 \sqrt{K}~\phi (x)]: .
\label{hb0}
\eea
 This dimerization contribution for NN exchange interaction has
originated from the XY interaction. The other two contribution of $J1$ dimerization are from XXZ 
anisotropy of NN exchange interaction and z-component of NNN
interaction. 
In recent past, there are considerable disagreement in
conclusions.  
Chen { $et~ al.$} \cite{chen} claimed that in the limit of small frustration the 
spin-Peierls dimerization
in NNN exchange interaction destabilizes the isotropic Heisenberg fixed point, leading 
to a new gapless Luttinger liquid phase with vanishing spin wave velocity. Our
study \cite{sar1} has found that spin-Peierls dimerization in NNN exchange interaction present an 
irrelevant perturbation in the regime of small frustration. An explicit
derivation of this $m=0$ plateau is
relegated in the appendix.
In this limit Parolla {\it et al.} \cite{cap} have also studied this model, 
and their conclusions are as the same as ours \cite{sar1}. So it is a time relevant problem, 
present author would like  
to see this limit of plateau phase more explicitly. Now we would like to discuss this situation
in the spinless fermions representation.

For the case of $J_1$ dimerization, we only consider the dispersion of 
free spinless fermions due to $J_1$, 
because the spin-Peierls perturbation will act only  
over the free spinless fermions due to $J_1$.  
The energy dispersion is $ {\epsilon}_k = - J_1~ cosk$, so in zero magnetic field, $k_F=\pm~\pi/2$. 
As a result wave vector ($Q= \pi$) of dimerized potential is commensurate
with the $2 k_F=\pi$ component of density wave and then it produce a gap 
in the excitation spectrum. 

But the explanation for absence of, gapped state for $J_2$ dimerization is not so obvious.
Only the intrinsic dimerization due to NNN exchange
interaction can drive a system from gapless, spin-fluid, state to gapped dimer order
state as one obtained from the Majumdar-Ghosh model \cite{maj}.
Numerically it has predicted \cite{pati} for isotropic ($\Delta = 1$) $J_1$-$J_2$ spin chain that
the system drives from spin-liquid state to dimer order state for $J_2 ~=0.241 J_1$.
In our theoretical calculations, we predict this transition for $J_2 ~=0.167 J_1$.
Parola $et~al.$ have studied, numerically, the effect of $\delta$ on $J_2$ dimerization,
there conclusion is negative, as we obtain analytically.   
It revealed from the variational study that the dimerization has only appreciable effect
for higher values of NNN \cite{sar1}.
It is clear from Fig. 2A of our study that the NNN exchange interaction helps to produce a
stable plateau phase for $J_1$  and $J_2 $ dimerization, if one only consider the umklapp term
due to the frustrating NNN exchange. 
Inset of Fig. 2A, of our study present the variation of critical ${(\frac{J_2}{J_1})}_c$ with the
XXZ anisotropy. 
Below the critical line system
display a short range anti-ferromagnetic order with gapless excitations by creating the
low lying spinon states and above the critical line it is the gapped dimer order state.  
It reveals that lower values of $\Delta$ 
shift more the critical value of ${(\frac{J_2}{J_1})}_c$ to change the 
phase from gapless spin-fluid phase to gapped dimer order phase. This is due to the
enhancement of quantum fluctuation of the fields for lower $\Delta$, {\it i.e.}, 
for the higher anisotropy.
\begin{figure} 
\includegraphics[scale=0.35,angle=0]{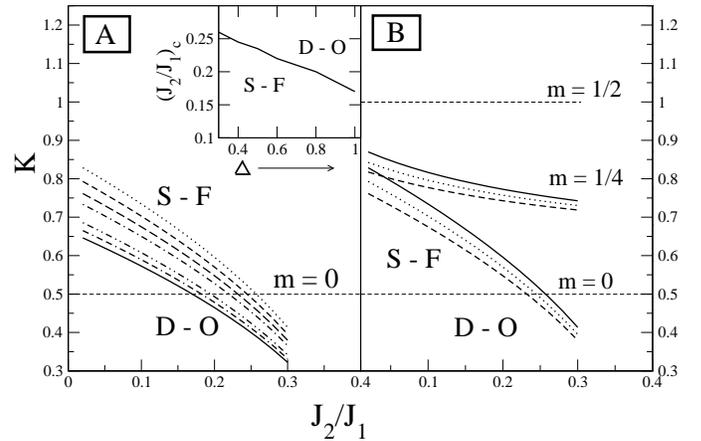} 
\caption{Luttinger liquid parameter ($K$) versus $\frac{J_2}{J_1}$ for different magnetization
plateaus. A constant dashed line at $K=0.5$ is for eye guide line, which separating 
the spin fluid and dimer order instability state. 
D-O: dimer order instability state, S-F: spin fluid Luttinger liquid phase.
{\bf A:} Here we only focus at $m=0$ magnetization plateau.
Uppermost (dotted) curve is for $\Delta =0.3$ and the lowermost (solid) curve is
for $\Delta =1$. The intermediate curves are for $\Delta = 0.4, 0.5, 0.6.0.8, 0.9$ respectively
from upper to lower one. Inset shows the separation between the spin fluid and
dimer order instability state by a critical line. Here we present the shift of the
${(\frac{J_2}{J_1})}_c$ with $\Delta$. {\bf B:} Here we present the 
curves for different magnetization plateaus. $m=1/2$ is independent of $\Delta$ and 
$\frac{J_2}{J_1}$ in contrast with $m=1/4 , 0$. The curves for $m=1/4 , 0$ plateaus are
for $\Delta = 0.4, 0.5, 0.6$ respectively from upper to lower one.} 
\label{Fig. 2}
\end{figure}

It is also clear from the Fig 2A and Fig. 2B that the Luttinger liquid
parameter ($K$) is the function of $\Delta$ and $J_2$ and applied magnetic field.
This dependence is  
modifying the anomalous scaling dimension of the physical field through $K$ 
and the commensurability
properties of spin operator. 
So we conclude that the effect of $J_2$ dimerization (intrinsic dimerization)
has pronounced effect near to the Heisenberg point ($\Delta=1$ and $K~=1/2$). 
Spin gap is maximum at this point.  
In the dimerized phase this excitations are triplet with a gap in the
excitation spectrum. One can conclude from the comparison of Fig. 2A (inset) of our study with the Fig.
4 of Ref. \cite{cap} 
that dimerization strength and XXZ anisotropy of  NN interaction varying reciprocally,
as it should be. Cross and Fisher have studied the spin-Peierls dimerization problem for
NN anti-ferromagnetic spin chain \cite{cross}. They have predicted that spin gap $\propto {(\delta J)}^{2/3}$ 
. We also retrieve the same result on that special 
limit of our general problem. Actually on that limit
spin gap is $\propto {(\delta J)}^{\frac{1}{2-K}}$ \cite{cross}. \\

\subsection{ 2.2 Calculations and Results for $m=\frac{1}{4}$ Magnetization Plateau:}

Here we discuss the occurence of a finite (nontrivial) magnetization plateau
We are considering the magnetization plateau at $ m=\frac{1}{4}$, it corresponds
$ k_F ~=~\pm \frac{\pi}{4}$. 
The effective Hamiltonian for $J_2$ dimerization become,
\bea
H_A & = & H_{0} ~-~ ( \frac{J_2 ~+~\delta}{2 {\pi}^2 {\alpha}^2} )~\int dx :cos[4 \sqrt{K}~\phi (x)]:
\nonum\\
&& + \frac{\Delta}{2 {\pi}^2 {\alpha}^2}~\int dx :sin[4 \sqrt{K}~\phi (x)]:. 
\label{hj2}
\eea
Apparently it appears from the general derivation of section (2), 
that the second and third terms of Eq. (\ref{hj2}) will be
absent due to the oscillatory nature of the integrand but this is not the case
when one consider the dimerized lattice. In dimerized lattice, reciprocal
lattice vector $G$ will change from $2 \pi$ to $\pi$ due to the change of the size of the 
unit cell. It become more clear, if one write these terms as
$ ~\int dx :cos[ (G-4 k_F)x ~+~ 4 \sqrt{K}~\phi (x)]:$.

Similarly one can write the effective Hamiltonian for $J_1$ dimerization:
\bea
H_B & = & H_{0} ~-~\frac{J_2}{2 {\pi}^2 {\alpha}^2}~\int dx :cos[4 \sqrt{K}~\phi (x)]:
\nonum\\
&& + \frac{\Delta~(1~+~\delta)}{2 {\pi}^2 {\alpha}^2}~\int dx :sin[4 \sqrt{K}~\phi (x)]:. 
\label{hj3}
\eea
The analytical structure of Hamiltonian (Eq. \ref{hj3} ) is the same as in Ref.6 for $J_1$
dimerization. 
The renormalization group equations for these type of interactions are \cite{tot,gog,gia2,kos,sub}.
\beq
\frac{dK}{dlnL}~=~ -4 {\pi}^2  K^2 {\delta}^2
\eeq 
\beq
\frac{d \delta}{dlnL}~=~ (2~-~4 K) \delta
\eeq 
It appears from these RG equations that to get a relevant perturbation, $K$ should be less than $1/2$.
It reveals from Fig. 3B that $K$ is exceeding the relevant value in our region of interest
to mature criteria for magnetization plateau. So the dimerization strength should exceed some critical
value (${\delta}_c$) to initiate the plateau phase.  
These two equations are the Kosterlitz-Thousless equation for the system in this limit. At the critical point,
system undergoes Kosterlitz-Thouless transition \cite{tot,gog,gia2,kos,sub}.
Here we are not interested to present the renormalization group flow diagram because it
has already discussed in Ref. \cite{tot}. The relation between the compactification radius ($R$) of Ref. \cite{tot} 
and the Luttinger liquid parameter ($K$) of our study is $K~=~R^2$.
As we have mentioned that this phase diagram has already explored explicitly
 \cite{tot,gog,gia2,kos,sub,suj2},
so we only present the relevant part of the phase diagram following Totsuka \cite{tot} and Tonegawa\cite{tone}:
since the system flows to the strong coupling (dimer-order) as the dimerization strength exceeds some critical
value (${\delta}_c$) initially, we have to guess the physics of this phase. We analyze the system in the
limit $\delta~ \rightarrow \pm \infty$ and $K~ \rightarrow 0$. In this limit we can expect that
the value of $\phi$ is pinned at one of the minima of $cos( 4 \sqrt{K} \phi)$.
This parameter dependent transition, from massless phase to massive phase, at $T=0$ is
the quantum phase transition \cite{sub}. This quantum phase transition is occurring at the beginning
and end of each magnetization plateau. 

\subsection{ 2.3 Calculations and Results for $m=\frac{1}{2}$ Magnetization Plateau:}

Now we discuss the saturation plateau at $m= \frac{1}{2}$ ($k_F ~=~0$).  
$K_F ~=~0$ implies that the band is empty and the dispersion is not linear, so the
validity of the continuum field theory is questionable. Values of the two 
Luttinger
liquid parameters, $v_0$ and $K$, are $0$ and $1$ respectively. 
It also implies that none of the sine-Gordon coupling terms become
relevant in this parameter
space.
Saturation plateaus are only appearing due to very high values of magnetic field.
In this plateau system is in ferromagnetic ground
state and restore the lattice translational symmetry. 
We think, this is the classical phase of the system.
Recently Cabra {\it et al} have studied 
the magnetization plateau for two leg ladder system with dimerized leg \cite{cab}. The configuration
of leg dimerization is different from our present study, Fig. 1. In this configuration,
each leg is dimerized by bond alternation and the two legs are connected by a rung coupling.
In their study relevant operator
arises from a combined effect of the interchain coupling and the dimerization along the chain.
This plateau is the manifestation of applied potential not due to the higher values of magnetic
field. This phase is the quantum phase of the system. 

Here we present the possible explanation for the absence of other fractionally
quantized magnetization plateaus (like $\frac{1}{3},\frac{1}{5}$ etc):
A carefull examination of Eq. 13 to Eq. 15 reveals that to get a nonoscillatory
contribution from  Hamiltonian
one has to be satisfied $4 k_F=G$ condition but this condition is not fulfilled for these
plateaus. The integrand of this sine-Gordon coupling terms contain an
oscillatory factor that leads to a vanishing contribution. The other criteria is
that non vanishing sine-Gordon coupling term should be relevant.
This incident is same for both $J_2$ and $J_1$ dimerized chain. 

\section{ 3. Discussions and conclusions:}
We have presented the generic feature of dimerized spin chain with NN and NNN exchange
interaction under a magnetic field. Different quantized magnetization 
plateaus have occurred. The occurrence of $m=0$ magnetization plateau of NN exchange is spontaneous
, whereas for NNN exchange interaction is not spontaneous.
We have also predicted a phase transition line, which separate gapless spin-fluid phase to 
gapped dimer order state as a function of $\frac{J_2}{J_1}$ and $\Delta$. We expect that this 
critical line is the same for $J_1$ and $J_2$ dimerization, if one only consider the dimerized
phase from intrinsic umklapp term. We also retrieve the result of Cross and Fisher \cite{cross}
in one special limit
of our general consideration. The basic origin of $m=\frac{1}{4}$ magnetization 
plateaus are the same for $J_1$ and $J_2$  
dimerization because a critical strength of dimerization is needed to promote the state into
a plateau phase. 
A critical strength of dimerized potential is necessary
for the occurrence of this plateau in presence of NNN interaction. One 
can expect the Kosterlitz-Thouless type of transition for the occurrence of this
plateau. This Kosterlitz-Thouless transition has
been  supported by the works of Totsuka \cite{tot}
and Tonegawa \cite{tone}.
Magnetization plateau at $m=\frac{1}{2}$ is the classical state of the system under high magnetic 
field. Cabra {\it et al.} \cite{cab} have 
predicted the $m=\frac{1}{2}$ magnetization plateau for zig-zag ladder
with different leg dimerization than our present problem. This plateau is the quantum phase 
of the system.  
We have explained the absence of other fractionally
quantized magnetization plateaus for our present problem.
Magnetic field dependence response for a zig-zag ladder with modulation in exchange
coupling have already studied by different groups numerically and they predict the
existence of non zero magnetization plateau \cite{umi}. 
A higher order commensurabilities have occured for nonzero magnetization plateaus. This transition
from commensurate gapped phase to gapless Luttinger liquid phase is the Mott-$\delta$ type of
transition.\\
 
Author would like to acknowledge Prof. T. Giamarchi for several extensive discussions during
the progress of this work and also Prof. B. I. Halperin, Prof. Dr. A. Honecker, and Prof. Diptiman Sen for partial
discussion. Dr. Abhishek Dhar and Dr. Nandan Pakhira are deserving acknowledgement for reading the
manuscript very critically.
Finally author would like to acknowledge 
Prof. A. M. Finkelstein and Dr. Yuval Oreg for many useful discussions 
on low dimensional many body systems and also for warm support.

\section{Appendix}

In this appendix we present the explicit bosonization derivation
for $m=0$ plateau
of $J_2$ dimerization.  Our starting Hamiltonian is  
\beq
H_{2\delta} = - J_2 ~\delta ~\sum_{n}~ (-1)^n ~\vec{S}_n \cdot \vec{S}_{n+2} ~.
\eeq
After Jordan-Wigner transformation ( Eq.\ref{jor}), Hamiltonian reduce to  
\bea
H_{2\delta} &=& - J_2 \delta \sum_n (-1)^n (\psi_{n+2}^{\dagger} 
\psi_n + {\rm h.c.}) (\psi_{n+1}^{\dagger} \psi_{n+1} - 1/2) \nonum \\ 
&& - J_2 \delta \sum_n (-1)^n (\psi_n^{\dagger} \psi_n - 1/2)
(\psi_{n+2}^{\dagger} \psi_{n+2} - 1/2) . \nonum \\
&&
\label{h2dfer}
\eea
We linearize the energy spectrum around the Fermi points $k_F~=\pm \frac{\pi}{2}$, and get
the Hamiltonian in a final form 
\bea
H_{2 \delta} &=& - J_2 \delta a^2 ~\sum_n ~ (-1)^n [ - ~(\rho_{n+1} -(-1)^n M_{n+1}) \times \nonum \\
&& ~~~~~~~~ (\rho_n + \rho_{n+2} + (-1)^n M_n + (-1)^n M_{n+2}) \nonum \\ 
&& + (\rho_n + (-1)^n M_{n}) ({\rho}_{n+2} + ~(-1)^n M_{n+2})].
\eea

During this derivation of $H_{2 \delta}$, we have used the following relations.
\bea
\rho (x) &=& ~-~ \frac{1}{\sqrt \pi} ~\partial_x \phi (x) ~, \nonum \\
M(x) &=& ~\frac{1}{\pi a} ~\cos (2  \phi (x)) ~.
\label{basic2}
\eea
Also,
we have used Taylor expressions such as
\beq
(R/L)(n+2) ~=~ (R/L)(n) ~+~ 2a(R'/L')(n) ~+~ 2a^2 (R''/L'')(n) ~+~ \cdots 
\eeq
to write
\bea
&& {(R/L)}^{\dagger}(n+2) {(R/L)}(n) ~+~ {(R/L)}^{\dagger}(n) {(R/L)}(n+2) \nonum \\
&& = {(R/L)}^{\dagger}(n+2) {(R/L)}(n+2) ~ +~ {(R/L)}^{\dagger}(n) {(R/L)}(n) + O(a^2) ~.
\eea
Where $R$ and $L$ are second quantized field of right and left moving fermions respectively.
The non-oscillatory contributions of the Hamiltonian for XY and Z component of 
interaction are 
\beq
H_{2\delta} (XY) = J_2 \delta a^2 \sum_n ~
 [\rho_{n+1} (M_n + M_{n+2}) - M_{n+1} (\rho_n + \rho_{n+2})].
\label{h2dbos}
\eeq 
\beq
H_{2\delta} (Z) = - J_2 \delta a^2 \sum_n ~[\rho_n M_{n+2} + M_n \rho_{n+2}].
\label{h2dbos}
\eeq
Now we perform an operator product expansion of the above Hamiltonian. In the
limit $z \rightarrow w$, we can use the following expansion \cite{nag},
\bea
\partial_z \phi (z) :e^{i \beta \phi (w)}: &=& - \frac{i\beta}{z-w} :e^{i 
\beta \phi (w)}: \nonum \\
&& + : \partial_z \phi (z) e^{i \beta \phi (z)}: ~
\label{ope}
\eea
For our case $\beta~=2$ and $K$ varies from less than $1/2$ to greater than
$1/2$ depending on $\Delta$ and $\frac{J_2}{J_1}$ ratio. Here the values of
$K$ are irrelevant because
the various relevant terms in $H_{2\delta}(XY)$
and $H_{2\delta} (Z)$ cancel each other due to the denominator, $z~-~w$
, which takes values $\pm a$, $\pm 2a$. 

Apart from this detail mathematical analysis one can also understand the absence
of this term from the following consideration: Here the applied dimerized potential
is staggered. The sine-Gordon coupling term due to this perturbation
is 
$$\frac{J_2 \delta}{2 {\pi}^2 {\alpha}^2}
 \int dx :cos[(\pi- 4 k_F)x~+~4 \sqrt{K}~\phi (x) ~-~4 k_F a]:.$$
So at $k_F~=~\frac{\pi}{2}$ contribute a oscillatory term in the integrand and it
leads to the vanishing contribution.

\end{document}